\documentclass[prb,preprint,showpacs,floatfix,superscriptaddress,epsf,psfig]{revtex4}
\usepackage{epsfig}
\begin{document}

\title{Controlling Band Gap in Silicene Monolayer Using External Electric Field}

\author{C. Kamal}

\affiliation{ Indus Synchrotrons Utilization Division, \\ Raja Ramanna Centre for Advanced Technology, Indore 452013, India }

\begin{abstract}
We study the geometric and electronic structures of silicene monolayer using density functional theory based calculations. The electronic structures of silicene show that it is a semi-metal and the charge carriers in silicene behave like massless Dirac-Fermions since it possesses linear dispersion around Dirac point. Our results show that the band gap in silicene monolayer can be opened up at Fermi level due to an external electric field by breaking the inversion symmetry. The presence of buckling in geometric structure of silicene plays an important role in breaking the inversion symmetry.  We also show that the band gap varies linearly with the strength of external electric field. Further, the value of band gap can be tuned over a wide range.

\end{abstract}

\maketitle

\section{Introduction}
There has been a lot of interest in silicene since it shows properties similar to those of graphene\cite{grap1,grap3,grap4,grap5}. For example, the theoretical studies on silicene show that the charge carriers in this two-dimensional material behave like massless Dirac-Fermions due to presence of linear dispersion around Fermi energy at a symmetry point K in the reciprocal lattice\cite{sili-ciraci1,sili-ciraci2}. Similar to its carbon counterpart - graphene, silicene is a potential candidate for applications in nanotechnology. The silicon based nanostructures have an important advantage that they are compatible with the existing semiconductor technology. Therefore, silicene and silicon nanoribbon have received much attention from both experimentalist and theoreticians\cite{sili-grown, sili-expt,sinr1,sinr2,sili-ciraci1,sili-ciraci2,sili2,sili3,refpap1,refpap2,refpap3,refpap5}. Recently, silicene has been epitaxially grown on a close-packed silver surface Ag(111)\cite{sili-grown}. Though graphene possesses many novel properties, its applications in nanoelectronic devices are limited due to its zero band gap and hence it is difficult to control the electrical conductivity of graphene. However, band gap in graphene may be introduced by chemical doping but chemical doping is uncontrollable and incompatible with device applications.  Hence, it is desirable to have band gap in materials in addition to their novel properties. 

In this letter, we demonstrate that silicene is one such a material in which the band gap can be opened up as well as varied over a wide range by applying external static electric field in a direction perpendicular to the plane of sheet. We use SIESTA package\cite{siesta1,siesta2,siesta3} for performing a fully self-consistent density functional theory (DFT) calculation by solving the standard Kohn-Sham (KS) equations. The KS orbitals are expanded using a linear combination of pseudoatomic orbitals proposed by Sankey and Niklewski\cite{sankey}. All our calculations have been carried out by using triple-zeta basis set with polarization function. The standard norm  conserving Troullier-Martins pseudopotentials\cite{tm} are utilized. For exchange-correlation potential generalized gradient approximation given by Perdew-Burke-Ernzerhof\cite{pbe} has been used. A cutoff of 400 Ry is used for the grid integration to represent the charge density and the mesh of k-points for Brillouin zone integrations is chosen to be 45$\times$45$\times$1. The convergence criteria for energy in SCF cycles is chosen to be 10$^{-6}$ eV. The structures are optimized by minimizing the forces on individual atoms (below 10$^{-2}$ eV/ $\AA$).  Super-cell geometry with a vacuum of 14 $\AA$ in the direction perpendicular to the sheet of silicene is used so that the interaction between adjacent layers is negligible.

\begin{figure}[ht]
\begin{center}
\includegraphics[width=8.5cm]{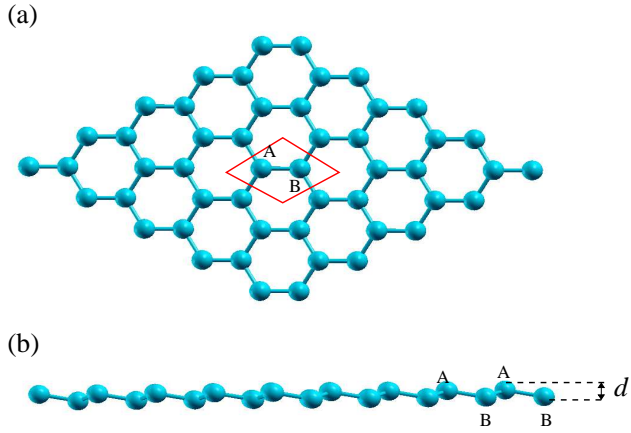}
 \caption{ The optimized structure of silicene monolayer. (a) Top and (b) side view of 5$\times$5 supercell. The lines in (a) represents the unitcell. The vertical distance between two Si atoms at sites A and B is represented by '$d$'.}
\end{center}
 \end{figure}
The optimized geometry of silicene is shown in Fig. 1. The unitcell has two Si atoms (A and B) and the space group is P3m1. We observed that the minimum energy structure of the silicene is low-buckled with the lattice constant of 3.903 $\AA$. The bond length and bond angles between the silicon atoms are 2.309 $\AA$ and 115.4$^\circ$ respectively.
The value of  bond angle in silicene lies in between those of sp$^2$ (120$^\circ$) and sp$^3$ (109.47$^\circ$) hybridized structures. This clearly shows that the hybridization in silicene is not purely sp$^2$ but a mixture of sp$^2$and sp$^3$. The buckling in silicene is due to weak $\pi$ - $\pi$ bond that exists between Si atoms since the Si-Si distance is much larger as compared to that in graphene (C-C =1.42 $\AA$).  The system increases its binding energy due to buckling by increasing the overlap between $\pi$ and $\sigma$ orbitals. The presence of buckling may also be explained by Jahn-Teller distortion. Our results on geometry match well with the previous calculations\cite{sili-ciraci1,sili-ciraci2,sili2,sili3,refpap1}. Here, we quantify the amount of buckling in terms of buckling length $d$ which is defined as the vertical distance between atoms at sites A and B in the unitcell. The value of $d$ in silicene is 0.5 $\AA$. In case of planar structure, as that of graphene, the value of $d$ is zero. The importance of $d$ in determining the electronic structure of silicene under the influence of external electric field will be described later. \\

\begin{figure}[ht]
\begin{center}
\includegraphics[width=8.5cm]{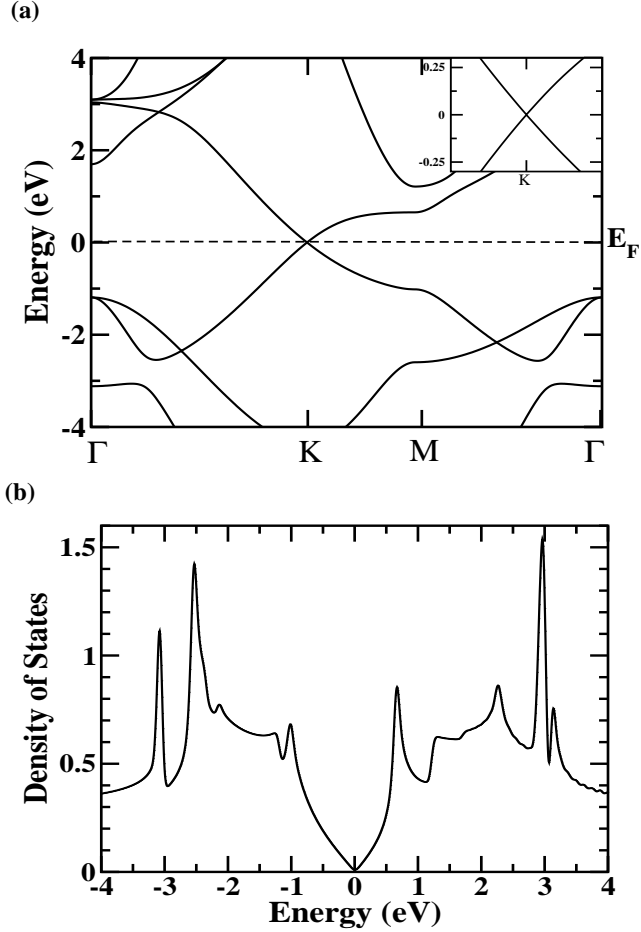}
 \caption{ (a) Band  structure and (b) density of states for optimized structures of silicene. The energy of bands are with respect to Fermi level.  }
\end{center}
 \end{figure}

The band structure along high symmetry points in Brillouin zone and density of states (DOS) of silicene are shown in Fig .2. The figure clearly indicates the semi-metallic behavior of silicene since the value of DOS at E$_F$ is zero and the conduction and valence band touch each other only at the symmetry K point. The energy levels and the contribution of density of states just below and above Fermi levels are mainly due to $\pi$ and $\pi^*$ orbitals. The dispersion around K point near the Fermi level is linear (see insert in Fig. 2 (a)). The point in E-k diagram where the conduction  and valence band touch each other at E$_F$ is called the Dirac point. The presence of linear dispersion indicates that the charge carriers near Dirac point behave like massless Dirac Fermions since the dynamics of these carriers obey relativistic Dirac-like equation. The relativistic Dirac-like Hamiltonian which describes the electronic structure of silicene around the Dirac point, similar to that of graphene\cite{oosting,wallace}, can be approximated as
\begin{eqnarray}
 \hat{H}= \left( {
\begin{array}{cc}  
\Delta & \hbar v_F (k_x-ik_y)  \\  
\hbar v_F (k_x+ik_y) & -\Delta  \\  
\end{array} } 
\right)
\end{eqnarray}
where $k$ and $v_F$ are momentum and Fermi velocity of charge carriers near Dirac point. The quantity $\Delta$ is the onsite energy difference between the Si atoms at sites A and B.  Due to the presence of inversion symmetry, the onsite energy difference $\Delta$ becomes zero, which leads to the linear dispersion around the Dirac point linear i.e. $E=\pm \hbar v_F k$.  Until now, the treatment for description of electronic structures of silicene is exactly similar to that of graphene. The band gap in graphene-like structure can be opened up if one can break the inversion symmetry and hence the value of  $\Delta$ becomes finite. Then, the dispersion around the Dirac point becomes
\begin{eqnarray}
 E=\pm \sqrt{\Delta^2+ (\hbar v_F k)^2}
\end{eqnarray}
In this case, the value of band gap opened is twice that of the onsite energy difference. It is possible to make $\Delta$ non-zero in graphene by  chemical doping which cause the local environment around the sites A and B to be different. As mentioned earlier, tuning the value of $\Delta$ by chemical doping is a difficult task. In the above mentioned discussion, the effect of spin-orbit coupling (SOC) is not included and very small gap of the order of $\mu$eV in graphene and about 1.55 meV in silicene may be opened up due to SOC\cite{grap-soc,sili-soc}. However, presence of small gaps may be useful in studying fundamental properties like quantum spin Hall effect and will not be useful in operating or controlling nanoelectronic devices made up of graphene/silicene at ambient conditions. In order to utilize many novel properties of graphene-like materials in nanodevices, it is desirable that the materials shall possess value of band gap more than the thermal energy kT at room temperature ($\sim$ 25 meV ).  \\

\begin{figure}[ht]
\begin{center}
\includegraphics[width=8.5cm]{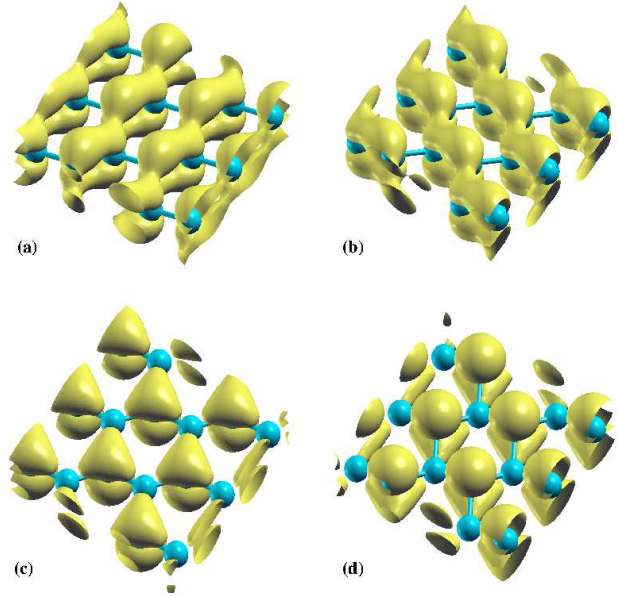}
\caption{Spatial distribution of highest occupied (without electric field (a) and with 5 V nm$^{-1}$ (c))  and lowest unoccupied states  (without electric field (b) and with 5 V nm$^{-1}$ (d)) }
\end{center}
 \end{figure}

In Fig. 3, we plot the spatial distribution of highest occupied and lowest unoccupied states with and without electric field. Both of these states lie at the symmetry K point in Brillouin zone. In the absence of external electric field, both the highest occupied  and lowest unoccupied states ( Fig. 3(a) and 3(b)) have same energy and also possess inversion symmetry. Hence the spatial distribution of these states above and below the silicene sheet looks similar. However, application of external electric field breaks the inversion symmetry in these states. This makes the spatial distribution of charges above and below the silicence sheet to be different which can be seen clearly from Fig. 3(c) and 3(d) respectively for highest occupied and lowest unoccupied states. This leads to the opening up of band gap in silicene. \\

\begin{figure}[ht]
\begin{center}
\includegraphics[width=8.5cm]{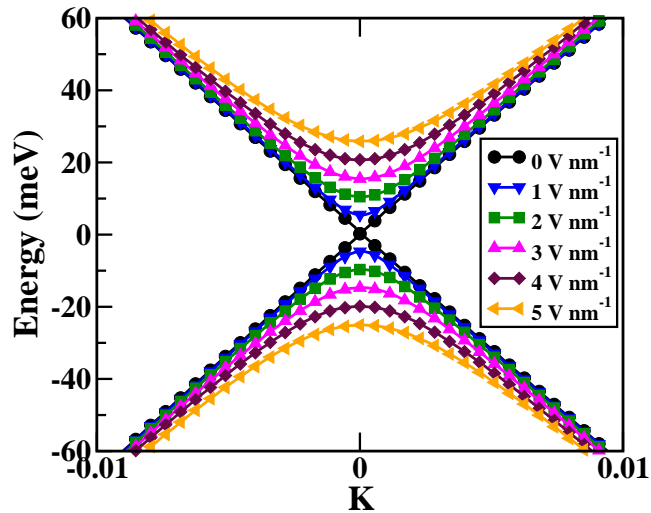}
\caption{ The band structure of silicene around Dirac point for different external electric field applied perpendicular to the plane of silicene sheet.  }
\end{center}
 \end{figure}

From our calculations on electronic structure of silicene, we observed that a band gap can be opened up at the Fermi level due to the influence of external electric field. The electric field is applied in a direction perpendicular to the plane of silicene sheet. The calculated band structures of silicene around Dirac point for various strengths of external electric field are shown in Fig. 4. For example, the value of band gap opened with electric filed of  5 volts per nanometer (V nm$^{-1}$) is  50.9 milli electron volts (meV). The band gap in silicene is opened due to breaking of inversion symmetry by the electric field since the potential seen by the atoms at the sites A and B are different. In this situation, the finite value of onsite energy difference arises due to the potential difference and hence, we can write 
 $\Delta = \alpha (V_A-V_B)$ where $\alpha$ and $V_A$ ($V_B$) are proportionality constant and the potential seen by the atom at the site A(B) respectively. When the field applied between the layers is constant around the sheet, as in the present case, the potential difference becomes $(V_A-V_B)= F $d$ $, where $F$ is strength of electric field and $d$ the distance between the atoms at the sites A and B along the field direction. The finite value of $d$ in silicene monolayer plays an important role in breaking symmetry in presence of electric field and cause opening of band gap. This result leads to an important advantage of silicene over graphene since there is no buckling in latter and hence it is not possible to open up a gap by applying external electric field. \\

\begin{figure}[ht]
\begin{center}
\includegraphics[width=8.5cm]{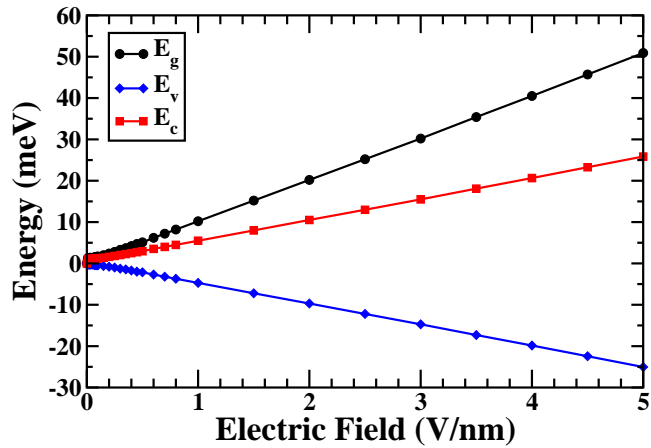}
\caption{The variation of band gap (E$_g$), valence (E$_v$) and conduction (E$_c$) band edges of silicene with the strength of external external field applied perpendicular to the plane of silicene sheet.}
\end{center}
 \end{figure}

We also observed from our DFT calculations that band gap can be tuned over wide range by varying strength of external electric field. The results of variation in band gap of silicene under influence of different strengths of electric field are summarized in given Fig. 5. It is seen from Fig. 5 that the band gap varies linearly with the external electric field. We also observed that both valence and conduction band edges move symmetrically away from Fermi level when the electric field is applied. The presence of  linear relationship between the band gap and electric field can be explained as follow.  As mentioned earlier, the value of band gap opened can be written as $E_g=2 \Delta = 2 \alpha (V_A-V_B) = 2 \alpha F $d$ = \alpha' F$, where $\alpha'$ is proportionality constant and it should be characteristic of the material. From Fig. 5, the value of the  $\alpha'$ is found to be 10.14 (meV per V/nm). It is remarkable that in silicene monolayer, the band gap can be tuned by simply varying the strength of external electric field.  Similar trends have also been observed for germanene monolayer and the results will be published elsewhere\cite{ck-germanene}.

In summary,  we have carried out abinitio DFT calculations to study geometric and electronic structure of silicene. Our results on the electronic structures of silicene monolayer show that it is a semi-metal and possesses linear dispersion around Dirac point similar to graphene and hence the charge carriers in silicene behave like massless Dirac-Fermions. We have shown that band gap in silicene monolayer can be opened up by applying external electric field. The presence of buckling in optimized structure of silicene plays an important role in breaking inversion symmetry. We also observe that the band gap varies linearly with the strength of electric field. Furthermore, our results predict that the band gap produced can be more than the thermal energy by applying electric field strength of few V/nm and hence, there is a possibility of using silicene monolayer in nanodevice even at room temperature.

Authors thank Dr. P. D. Gupta and Dr. S. K. Deb for encouragement and support. Thanks also to Dr. Aparna Chakrabarti, Dr. Arup Banerjee and Dr. J. Jayabalan for critical reading of the manuscript. The support and help of Mr. P. Thander and the scientific computing group, Computer Centre, RRCAT is acknowledged.

\end{document}